# Imidazolium Ionic Liquid Mediates Black Phosphorus Exfoliation while Preventing Phosphorene Decomposition

Vitaly V. Chaban,[1] Eudes Eterno Fileti,[1] and Oleg V. Prezhdo[2]

(1) Instituto de Ciência e Tecnologia, Universidade Federal de São Paulo, 12247-014, São José dos Campos, SP, Brazil.

(2) Department of Chemistry, University of Southern California, Los Angeles, CA 90089, USA.

**Abstract**. Forthcoming applications in electronics and optoelectronics make phosphorene a subject of vigorous research efforts. Solvent-assisted exfoliation of phosphorene promises affordable delivery in industrial quantities for future applications. We demonstrate, using equilibrium, steered and umbrella sampling molecular dynamics, that the 1-ethyl-3-methylimidazolium tetrafluoroborate [EMIM][BF$_4$] ionic liquid is an excellent solvent for phosphorene exfoliation. The presence of both hydrophobic and hydrophilic moieties, as well as substantial shear viscosity, allows [EMIM][BF$_4$] simultaneously to facilitate separation of phosphorene sheets and to protect them from getting in direct contact with moisture and oxygen. The exfoliation thermodynamics is moderately unfavorable, indicating that an external stimulus is necessary. Unexpectedly, [EMIM][BF$_4$] does not coordinates phosphorene by π-electron stacking with the imidazole ring. Instead, the solvation proceeds via hydrophobic side chains, while polar imidazole rings form an electrostatically stabilized protective layer. The simulations suggest that further efforts in solvent engineering for phosphorene exfoliation should concentrate on use of weakly coordinating ions and grafting groups that promote stronger dispersion interactions, and on elongation of nonpolar chains.

**TOC**

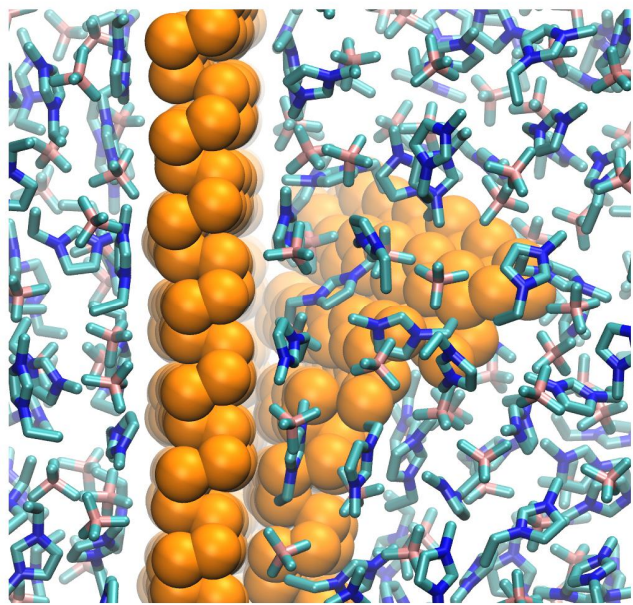 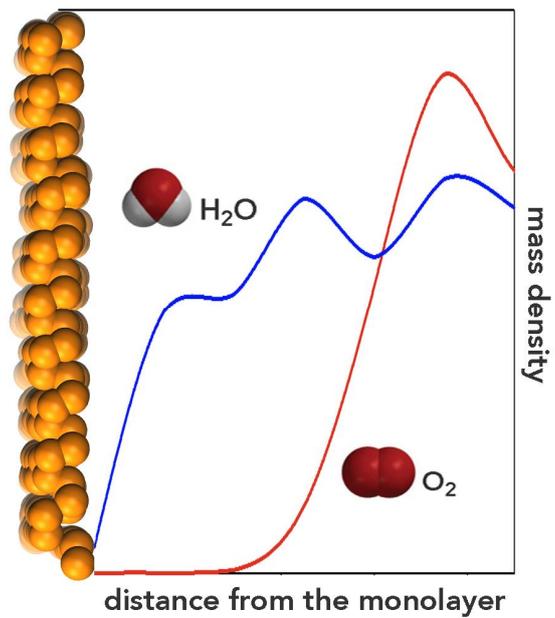

**Introduction**

During the last decade, investigation of 2D materials has become one of the hottest areas of nanoscience. These novel 2D systems – graphene, molybdenum disulfide, boron nitride, phosphorene, etc – exhibit unusual electronic and physicochemical properties, making them candidates for a variety of applications, particularly in electronics and optoelectronics,[1-11] as well as solar cells, lithium-ion batteries, photodetectors, transistors, and even gas sensors.[12-24] An atomically thin, covalently linked sheet of black phosphorus (BLP), phosphorene represent a newer member of this family.[25] The number of layers plays a crucial role, as it allow to tune the semiconductor band gap, from 0.3 eV in black BLP to 2.2 eV in a single-layer phosphorene.[26] A significant effect of the number of layers suggests a strong interlayer coupling in BLP. Furthermore, both the field-effect mobility and the current on/off ratio depend crucially on the number of layers, offering an easy way to tune these physical properties. Monolayers are kept together in BLP by relatively weak van der Waals interlayer forces, in the same way as graphene monolayers form graphite. In principle, BLP can be exfoliated by mechanical cleavage to produce a mixture of oligophosphorenes.[27] Since Raman scattering characteristics are sensitive to the thickness of oligophosphorene, determination of structure of an exfoliated sample is possible. It must be noted that Raman measurements are generally considered quite challenging, particularly in terms of the poor signal-to-noise, requirements for high resolution detectors, and the need to do the experiments in an inert atmosphere. Since the demand for phosphorene is large, mechanical cleavage, in its present formulation, is unable to produce sufficient quantities of phosphorene, calling for urgent development of alternative techniques.

Solvent-assisted exfoliation has been successfully used to obtain other 2D materials, such as graphene and boron nitride, in substantial amounts.[28-30] While spontaneous exfoliation is probably impossible, centrifugation and sonication of BLP crystals in suitable liquids can be used to overcome interlayer forces and produce mono- and several-layer phosphorenes.[25, 31, 32] An exfoliation leading to formation of a uniform dispersion of phosphorene may be possible in N-methyl-2-pyrrolidone, as described recently.[33] However, the demonstrated production yield is modest yet. The newly obtained phosphorene is chemically unstable in many other molecular solvents, such as water, thereby hampering prospective applications. In drastic contrast with other 2D materials, phosphorene stability is undermined in the presence of oxygen. A suitable exfoliation solvent will minimize oxidation by forming a stable protective solvation shell.

The development of promising solvents for phosphorene exfoliation is underway. Theoretical methods in chemistry may be useful for compound screening, since they allow to directly obtain phosphorene-liquid interaction energies based on appropriate structural models and simulated thermal motion. Blankschtein and coworkers[34] employed classical molecular dynamics (MD) simulations to rank dimethyl sulfoxide, dimethyl formamide, isopropyl alcohol, N-methyl-2-pyrrolidone and N-cyclohexyl-2-pyrrolidone for phosphorene exfoliation. Planar molecules, such as N-methyl-2-pyrrolidone, were found to orient parallel to the solid/liquid interface. The importance of quick solvent intercalation between phosphorene layers for efficient exfoliation was demonstrated. Observed, for instance, in dimethyl sulfoxide, strong solvent-solvent interactions in the first solvation shell surrounding monophosphorene were found useful to prevent immediate backward association of the sheets. While the exfoliation free energy computed for the molecular solvents was appropriate for applications, it remained to be seen whether these solvents could protect phosphorene surfaces from the reactive species. Based on the recent investigation, chemical instability presents challenges for BLP liquid-phase exfoliation, in particular, surfactant micelles in aqueous solution do not fully exclude solvated oxygen and water, allowing potential reactions between BLP and these oxidizing agents.[35]

Recently, Zhao and co-workers[32] demonstrated the utility of imidazolium room-temperature ionic liquids (RTILs) for phosphorene exfoliation. A facile, large-scale, and environmentally friendly liquid-exfoliation method was developed to produce stable dispersions of mono- and oligophosphorenes, with concentrations up to 0.95 g L$^{-1}$. The prepared suspensions maintained stability for one month without sedimentation and aggregation in the presence of air. The experiments showed 1-hydroxyethyl-3-methylimidazolium trifluoromethansulfonate to be the most efficient imidazolium-based RTIL. The systems obtained after the exfoliation could be seen as supersaturated, and thus metastable, solutions of phosphorene. Preservation of these solutions over macroscopic times constituted a very important achievement. Molecular solvents maintain systematically less dilute dispersions of the BLP nanoflakes. For instance, the concentrations reported in DMSO and DMF are below 0.01 g L$^{-1}$ after centrifugation and sonication.[36] N-methyl-2-pyrrolidone was shown to be much more successful, ~0.4 g L$^{-1}$,[37] but still less productive than the imidazolium-based RTIL, as discussed above.

We report the first theoretical study of phosphorene exfoliation in an ionic liquid. We selected the imidazolium-based RTIL due to availability of certain experimental data favoring further investigation of this family of solvents for the phosphorene exfoliation.[32] The study demonstrates that 1-ethyl-3-methylimidazolium tetrafluoroborate [EMIM][BF$_4$] provides a good choice of solvent to assist with exfoliation of phosphorene. [EMIM][BF$_4$] both protects phosphorene sheets against exposure to the reactive species and lowers significantly the free energy barrier to the exfoliation. The solvent assisted exfoliation is favored enthalpically, but is inhibited entropically. The mutual orientation of phosphorene sheets is never parallel after the exfoliation has been initiated. [EMIM][BF$_4$] is a moderately viscous RTIL, containing both polar and nonpolar moieties. The relatively high viscosity of [EMIM][BF$_4$], 36.9 mPa at room conditions,[38] is useful for slowing down the backward association of the phosphorene sheets, assisting to maintain a long-lived metastable exfoliated state. The surface tension of [EMIM][BF$_4$] is 53 mN m$^{-1}$ at 298 K.[39] In turn, the surface energy of the bilayer phosphorene, estimated theoretically from steered exfoliation, is ~58.6 mJ m$^2$.[40] Similarity of the surface energies of the solute and the solvent favors BLP exfoliation thermodynamically. Since phosphorene is virtually nonpolar, it is coordinated by the hydrocarbon chains of EMIM$^+$, whereas the imidazole rings form an electrostatically stabilized solvation shell. The latter prevents oxidizing species from approaching the phosphorene surface and, therefore, enhances its chemical stability. Non-polar oxygen is poorly soluble in [EMIM][BF$_4$] confirmed both by our simulations and previous experiments.[41] Therefore, oxidation of the exfoliated phosphorene by O$_2$ oxygen is impossible. Water can approach phosphorene sheets, however, the water density near phosphorene is an order of magnitude smaller than in bulk [EMIM][BF$_4$]. The study provides specific suggestions for further efforts in solvent engineering for phosphorene exfoliation. Elongation of the hydrophobic side chains and attachment of low-polarity groups containing polarizable atoms should improve the RTIL performance. Weakening of cation-anion interactions[42] should also have a positive effect. The predictions and design principles can be tested directly by exfoliation experiments.

**Results and Discussion**

The atomistic modeling of the phosphorene exfoliation process is carried out using a combination of equilibrium, steered, and umbrella-sampling MD simulations, as detailed in the Methodology section. Figure 1 exemplifies the equilibrated geometries of the simulated MD systems. The structural properties and pairwise interaction energies are obtained from equilibrium MD, while the thermodynamics of exfoliation are derived by umbrella-sampling MD along the exfoliation reaction coordinate. The solvation free energy is estimated by step-wise alchemical decoupling of the solute (phosphorene) from the solvent environment (RTIL).

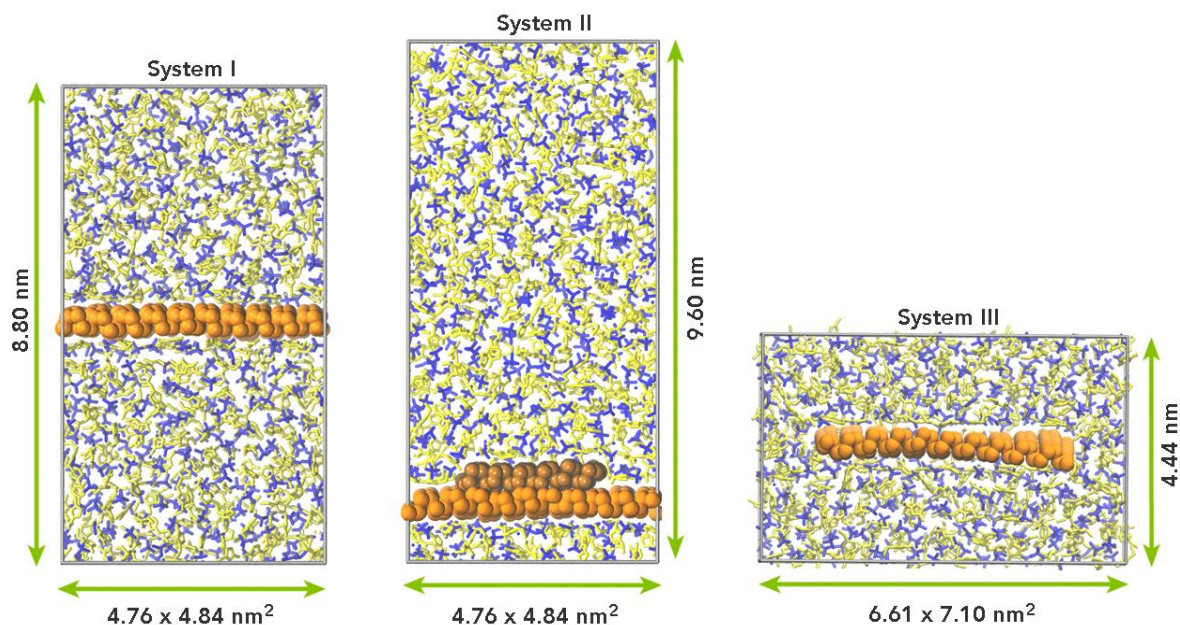

**Figure 1:** (Left) Simulated unit cell containing an infinite phosphorene sheet, system I. (Center) PMF calculation setup, system II. The distance between the centers-of-mass of the sheets is the reaction coordinate. (Right) Solvated phosphorene quantum dot, system III. Phosphorene quantum dots of sizes 2.0 nm × 2.5 nm and 4.2 nm × 4.8 nm were constructed for PMF and free energy calculations, respectively. Phosphorus is orange, cations are yellow, anions are blue.

Figure 2 illustrates the exfoliation process simulated using steered MD. One of the sheets (infinite phosphorene) is restrained, whereas the center-of-mass of the other sheet is forced to move in the normal direction at the rate of 0.001 nm ps$^{-1}$. The molecular configuration at the 1.4 nm center-of-mass separation shows flexibility of phosphorene and its resistance to separation. Note that the mutual orientation of phosphorene sheets is never parallel after the exfoliation has been initiated.

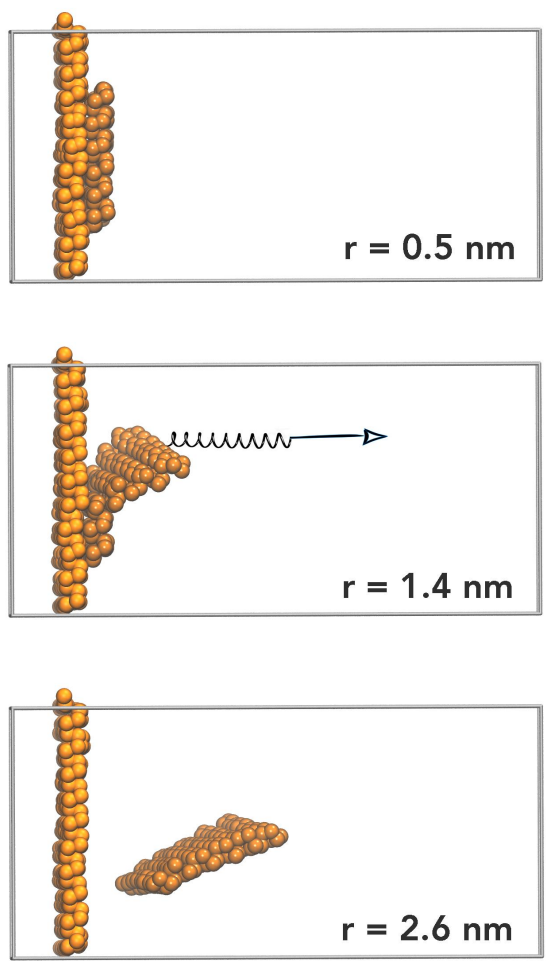

**Figure 2**: Three molecular configurations obtained in the course of steered MD simulations. The molecular configurations were subsequently used as starting points for the umbrella-sampling MD simulations. The designated distances refer to the centers-of-mass of the separated sheets. Ions are omitted for clarity.

The EMIM$^+$ cation approaches the phosphorene surface rather tightly (Figure 3, left). The first peak at ~0.3-0.4 nm corresponds predominantly to the side ethyl chain. The second peak at ~0.6 nm arises due to the imidazole ring and is accompanied by the anion's peak. The polar parts of [EMIM][BF$_4$] are also able to approach the surface of phosphorene sporadically, but those peaks are always smaller, as compared to the more distant peaks, ~0.6 nm, for the same atom pairs. Decomposition of the ions into the interaction centers (Figure 3, right) confirms the conclusions derived. Since water, which is often present in real-world [EMIM][BF$_4$] in trace quantities, binds to the imidazole ring, in particular to the acidic hydrogen atom, it would be located relatively far, ~0.6 nm, from the phosphorene surface and, therefore, the probability of the phosphorene decomposition reaction is initiation is insignificant.

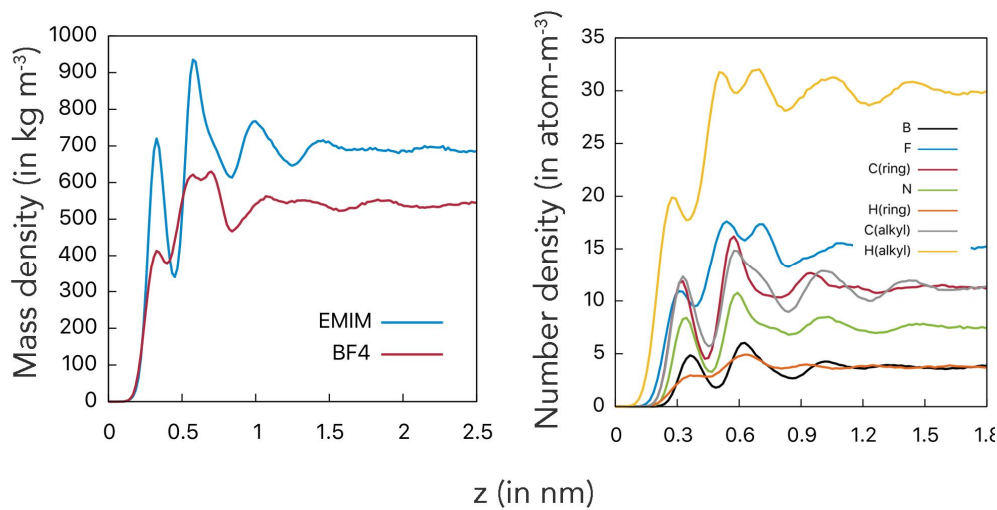

**Figure 3**: Mass and number density distributions calculated for phosphorene and [EMIM][BF$_4$]: (left) mass density profiles of EMIM$^+$ and BF$_4^-$; (right) number density profiles of the interaction centers belonging to [EMIM][BF$_4$].

Which interaction center of [EMIM][BF$_4$] brings the largest enthalpic gain? Figure 4 suggests that such centers are a nitrogen of the imidazole ring and a carbon of the imidazole side chain. This result is fairly intuitive, since solvation of phosphorene is driven by van der Waals interactions. The boron atom of the anion also brings a substantial energetic contribution, although the atom does not approach phosphorene directly. Analysis of pairwise interactions is useful to get an idea which atoms within the solvent particle are most useful to dissolve a given solute particle. There is no obvious correlation of pairwise solute-solvent interaction energies (Figure 4) and structure distributions (Figure 3). This fact suggests that solvation of phosphorene depends crucially on the solvent-solvent interactions, whose effect is accounted for in the free energy calculations by effective solute decoupling.

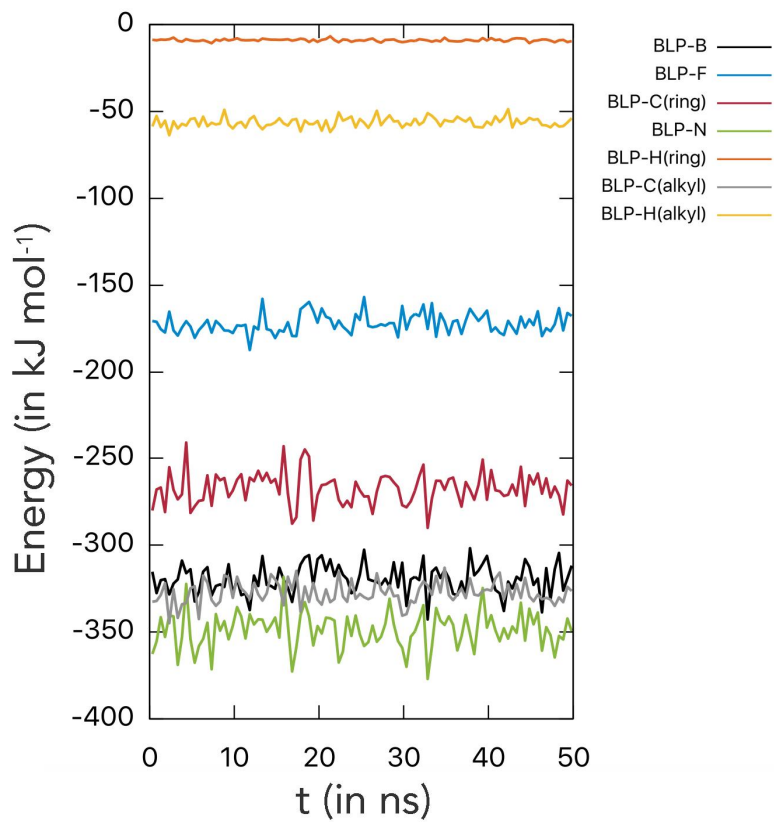

**Figure 4**: Decomposition of pairwise interaction energies between phosphorene and [EMIM][BF$_4$] into the contributions of the individual interaction sites. The energies are normalized with respect to the number of each interaction center in the ion pair.

We provide exfoliation PMFs both in the solvent and in vacuum to identify the impact of the solvent (Figure 5). These PMFs represent the cumulative reversible work that is required to separate two phosphorene monolayers by moving them in the opposite directions. The separation coordinate is defined by the distance between the centers-of-mass. The PMF curves are smooth. Unlike Blankschtein and co-workers, who studied molecular solvents,[34] we did not observe statistically meaningful inflection points. It is unclear whether the difference arises due to a different type of simulated liquid or due to a slightly different pulling scheme. Instead of pulling the edge atom of phosphorene, we gradually increased the distance between the centers-of-mass. Further investigation is required to identify which pulling scheme mimics real exfoliation most closely.

The Gibbs free energy penalty to exfoliation is reduced by ~500 kJ mol$^{-1}$ in the presence of [EMIM][BF$_4$], as compared to vacuum. [EMIM][BF$_4$] stabilizing the increase of the phosphorene surface area. As the number of monolayers grows in the simulated system, the van der Waals phosphorene-phosphorene attraction is replaced by the phosphorene-RTIL attraction. Due to a high density of phosphorus atoms in the phosphorene monolayer, the integral attraction energy is rather high, as revealed by a favorable solvation enthalpy, -11 kJ mol$^{-1}$ per phosphorus atom.

Blankschtein and co-workers[34] also concluded that solvents have a positive influence on the energetics of phosphorene exfoliation. The molecular liquids investigated in that work, dimethyl sulfoxide, dimethylformamide, isopropyl alcohol, N-methyl-2-pyrrolidone and N-cyclohexyl-2-pyrrolidone, result in similar exfoliation PMFs, with total free energy penalties amounting to over 300 kJ mol$^{-1}$. Note the PMFs reported by Blankschtein and co-workers did not reach the plateau, which would mean a complete exfoliation. Therefore, point-by-point comparison of the exfoliation works in the molecular solvents is not possible.

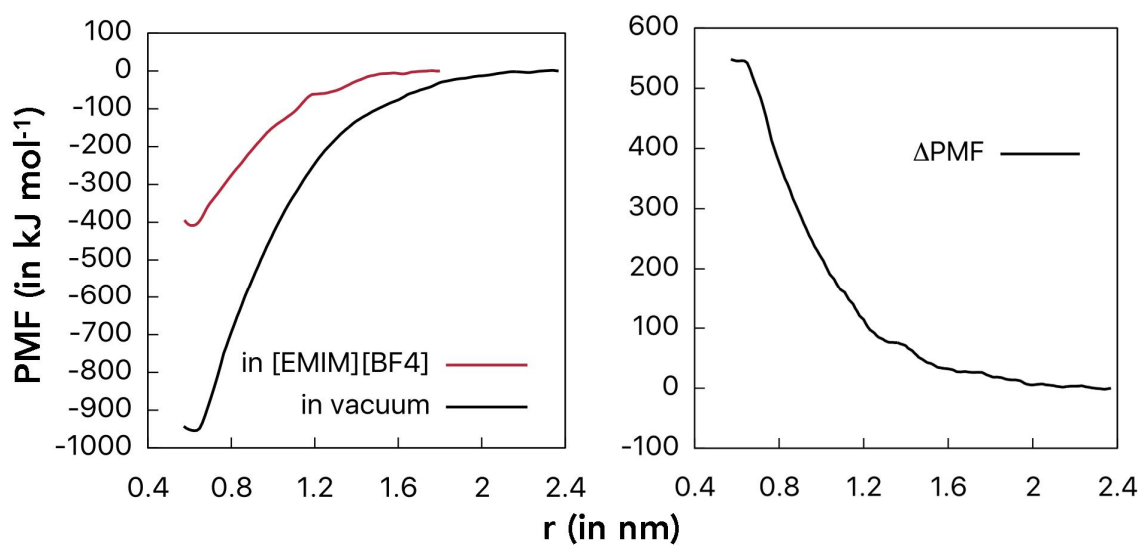

**Figure 5:** (Left) Potentials of mean force characterizing phosphorene exfoliation in [EMIM][BF$_4$] and vacuum. (Right) Difference between the two potentials. PMF=0 stands for the fully separated phosphorene monolayers. PMFs normalized per unit area (1 nm$^2$) of phosphorene are provided in Figure S2 to simplify point-by-point comparison with future phosphorene exfoliation studies.

The exfoliation PMF is in concordance with the free energy of solvation, i.e. transfer of monolayer phosphorene from vacuum to the fully solvated state in [EMIM][BF$_4$]. Solvation is favored enthalpically, -5.7 MJ mol$^{-1}$, but is prohibited entropically, +9.4 MJ mol$^{-1}$. Negative enthalpy implies that phosphorene gets covered rather efficiently by a solvation shell of RTIL. Due to a high density of phosphorus atoms on the phosphorene surface and their dispersion attraction to the nitrogen and carbon atoms of EMIM$^+$, the resulting enthalpy appears more negative than that of the solute and the solvent separately. Phosphorene represents a nanoscale particle, which decreases conformational freedom of the solvent. [EMIM][BF$_4$] is forced to adopt specific conformations to cover the monolayer and, at the same time, to maintain ionic interactions with the bulk RTIL. The resulting entropic penalty beats the enthalpic gain and leads to a positive Gibbs free energy of solvation, +3.7 MJ mol$^{-1}$. An interplay of enthalpy and entropy in phosphorene qualitatively resembles the previously discussed case of graphene.[29] The thermodynamic potentials depend uniformly on the size of the phosphorene sheet. To compare different monolayer sizes, it is necessary to normalize the potentials by the mole of phosphorus atoms or by the unit area, using parameters listed in Table 1. The dominance of entropic and dispersion contributions to the solvation properties of the 2D materials was recently highlighted.[43]

It is known that exfoliated phosphorene is reactive. Its prompt decomposition can be fostered by such omnipresent agents as moisture and molecular oxygen. A properly selected solvent should cover phosphorene monolayers to hamper access of reactive molecules to phosphorene surfaces. Many RTILs are viscous, and therefore, should block or, at least, slow down diffusion of reactive molecules. To further check suitability of [EMIM][BF$_4$] to protect phosphorene from reactive species, we added O$_2$ (system IV) and water (system V), and thoroughly equilibrated the MD systems. The spatial distributions of $O_2$ and water are provided in Figure 6. Non-polar oxygen is rather insoluble in [EMIM][BF$_4$]. Thus, no direct contact of $O_2$ and phosphorene was observed, and consequently, oxidation of the exfoliated phosphorene by oxygen is impossible. In the case of water, the distribution needs a detailed interpretation. [EMIM][BF$_4$] and water exhibit high mutual miscibility, and therefore, water molecules readily penetrate the bulk phase of [EMIM][BF$_4$]. However, phosphorene is predominantly coordinated by the non-polar moiety of EMIM$^+$, whereas water molecules coordinate the imidazole ring, particularly, the acidic hydrogen atom, C(2)-H. Due to this feature, water does not directly approach phosphorene, although it does exists in phosphorene vicinity. For instance, the water density near phosphorene amounts to 0.2 kg m$^{-3}$, which is an order of magnitude smaller than water density in bulk [EMIM][BF$_4$] (Figure 6). This separation of water from phosphorene by [EMIM][BF$_4$] is promising compared to polar molecular solvents, e.g. DMSO, whose molecule is small and spherical being unable to keep water molecules far from phosphorene. Note that we simulated an infinite phosphorene sheet of ideal geometry. However, it is possible that the phosphorene decomposition is initiated at its edges, and presumably defect sites, whereas the decomposition at the pristine basal surface is slow or negligible. Further simulations using reactive force fields (non-existing yet) would be desirable to shed light on the resistance of the exfoliated phosphorene to moisture and molecular oxygen.

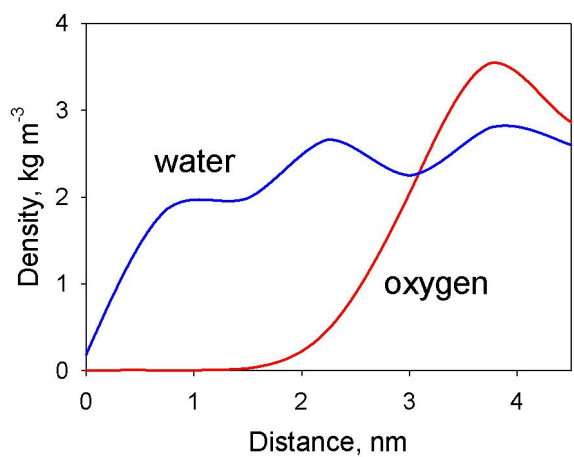

**Figure 6**. Mass densities of molecular oxygen and water as functions of distance from the phosphorene monolayer.

`

To recapitulate, we have reported the first theoretical investigation of phosphorene exfoliation in an ionic liquid. Equilibrium, steered, and umbrella-sampling MD simulations have been conducted to characterize the structure and thermodynamics of phosphorene solvation, the exfoliation potential of mean force, and the ability of the [EMIM][BF$_4$] film to protect phosphorene against water and oxygen.

The Gibbs free energy change corresponding to complete phosphorene exfoliation in [EMIM][BF$_4$] has been found somewhat positive, +400 kJ mol$^{-1}$ for the 5 nm$^2$ phosphorene monolayer, or +80 kJ mol$^{-1}$ nm$^{-2}$. The result is comparable to the previously studied exfoliation of graphene in the same RTIL, +70 kJ mol$^{-1}$ nm$^{-2}$.[29] Our result also agrees qualitatively with the data of Blankschtein and co-workers,[34] who have investigated several molecular liquids for phosphorene exfoliation. We argue that, unlike molecular solvents, [EMIM][BF$_4$] offers a more efficient protection against moisture and oxygen due to its higher shear viscosity and amphiphilic interactions.

While a promising performance of [EMIM][BF$_4$] for exfoliation and protection of phosphorene has been demonstrated, we anticipate further efforts in solvent design to proceed in the following directions. Since phosphorene is coordinated by the hydrophobic side chains, elongation of these chains may improve the RTIL performance. For instance, successful phosphorene exfoliation in N-cyclohexyl-2-pyrrolidone, in which cyclohexyl is an analogue of the hydrophobic chain, was recently reported.[44] The authors discuss production in large quantities by solvent-assisted exfoliation at room conditions and point out a surprising stability of the produced phosphorene. Attachment of low-polarity groups containing atoms with higher dispersion constants, e.g. sulfur, chlorine, phosphorus, to the side chains may also be useful to maximize the solute-solvent dispersion interactions. Weakening of cation-anion interactions has a positive effect on the solvation properties of RTILs.[42] In this context, weakly coordinating ions, e.g. bis(trifluoromethanesulfonyl)imide, tetrabutylammonium, tricyanomethanide and others, deserve attention.[45-47]

The results reported in this work inspire more proactive investigation and engineering of ionic liquids as phosphorene exfoliation media. Experimental validation of our predictions and design principles is rather straightforward by direct exfoliation experiments. Such experiments would be quite important for further progress in the field.

**Methodology**

Three MD setups were implemented (Table 1). First, an infinite phosphorene sheet was immersed in liquid [EMIM][BF$_4$] in the 4.76 nm × 4.84 nm × 8.80 nm unit cell (system I). This system was used to derive structure and thermodynamics of solvated phosphorene based on equilibrium constant-temperature constant-pressure MD simulations at 400 K and 1 bar. Second, two phosphorene sheets, 4.8 nm × 4.8 nm and 2.0 nm × 2.5 nm, were immersed in liquid [EMIM][BF$_4$] in the 4.76 nm × 4.84 nm × 8.80 nm simulation cell (system II). This box was used to obtain the exfoliation potential of mean force via umbrella sampling simulations. Third, the 4.2 nm × 4.8 nm phosphorene sheet was fully solvated in the 6.61 nm × 7.10 nm × 4.44 nm simulation cell (system III). Consequent positions of phosphorene along the exfoliation coordinate were restrained by applying potential $E = k \times (r - r_0)^2$, where $k = 2000$ kJ mol$^{-1}$ nm$^{-2}$, to the center-of-mass of the phosphorene edge. This system was used to derive solvation free energy via step-wise decoupling of the solute from the solvent. Ten oxygen molecules (system IV) and ten water molecules (system V) were added to phosphorene solvated in liquid [EMIM][BF$_4$] to compute structure correlations of these molecules and the phosphorene monolayer protected by the solvation shell of [EMIM][BF$_4$].

Phosphorene was simulated as a non-polarizable assembly of phosphorus atoms linked by harmonic potentials for bonds, angles, and dihedrals. The equilibrium intra-layer phosphorus-phosphorus distance is 0.222 nm, whereas the inter-layer distance is 0.225 nm. The equilibrium intra-layer angles are 102°, whereas inter-layer angles are 97°.[48] The Lennard-Jones (12,6) parameters for phosphorus atoms ($\sigma = 0.333$ nm, $\varepsilon = 2.092$ kJ mol$^{-1}$) were recently successfully used by Blankschtein and co-workers.[34] [EMIM][BF$_4$] was represented by the in-home force field (FF).[34] This FF provides correct thermodynamic and transport properties in a wide temperature range. The Lorentz-Berthelot combination rules were used to quantify phosphorene-RTIL interactions. Electrostatic interactions at the distances beyond 1.3 nm were accounted for by the Particle-Ewald-Mesh method.[49] The shifted force method was used to smoothly decrease Lennard-Jones forces to zero from 1.1 to 1.2 nm.

Constant temperature was maintained by the velocity rescaling thermostat with a relaxation time of 0.5 ps.[50] Simulations at the elevated temperature, 400 K, were used to accelerate sampling and avoid metastable states. Constant pressure, 1 bar, was maintained by the Parrinello-Rahman barostat[51] with the relaxation time of 4.0 ps and compressibility constant of $4.5 \times 10^{-5}$ bar$^{-1}$. The system was equilibrated during 5.0 ns. The equilibrium properties were sampled based on 50 ns long production runs. The integration time-step of 2.0 fs was used. Coordinates and intermediate thermodynamic quantities were saved every 10 ps.

The Bennett Acceptance Ratio method was employed to obtain solvation free energy (FE).[52] Phosphorene was decoupled from RTIL using 21 sequential states. After equilibrium was achieved at each state, the free energy contribution was averaged over a 10 ns long MD trajectory. The soft-core repulsion term for the LJ interactions was applied. Electrostatic and Lennard-Jones solute-solvent interactions were decoupled separately. Enthalpic component of the solvation free energy was taken as difference between total potential energy of phosphorene in vacuum and in [EMIM][BF$_4$]. The entropic component was determined directly, $\Delta G = \Delta H - T\Delta S$.

Umbrella sampling with weighted histogram analysis was used to derive PMFs. Phosphorene was pulled along the z-axis, normal to the phosphorene sheet, to generate 28 configurations ($\Delta z = 0.05$ nm as measured between the centers-of-mass of the sheets) in vacuum and in [EMIM][BF$_4$]. 8 additional configurations were sampled between z = 1.0 and z = 1.4 nm, to obtain more accurate energies. Each configuration was sampled during 11 ns, whereas the first nanosecond of each such simulation was disregarded as equilibration.

GROMACS 5 was employed to conduct MD simulations.[53] VMD 1.9.2 was used to prepare molecular artwork.[54] PACKMOL was used to prepare initial configurations.[55]

Table 1: Atomistic compositions of the investigated systems and their sampling times.

| Property | System I | System II | System III | System IV | System V |
|---|---|---|---|---|---|
| # interaction sites | 17 800 | 18 004 | 17 800 | 8940 | 8950 |
| # ion pairs | 720 | 720 | 720 | 350 | 350 |
| # phosphorus atoms | 520 | 724 | 520 | 520 | 520 |
| Phosphorene surface area, nm$^2$ | 46.0 | 46.0 (10.0)* | 46.0 | 46.0 | 46.0 |
| # independent simulations | 1 | 21 | 36 | 1 | 1 |
| Total equilibration time, ns | 5 | 21 | 36 | 10 | 10 |
| Total sampling time, ns | 50 | 210 | 360 | 40 | 40 |

*The area of the smaller sheet is given in parentheses.

**SUPPLEMENTARY INFORMATION**

Figure S1 exemplifies estimation of standard errors in the PMF calculations. Table S1 lists thermodynamic potentials of the phosphorene solvation normalized with respect to different parameters of the simulated phosphorene sheets. Figure S2 provides exfoliation PMFs normalized per unit area of the simulated phosphorene sheets.

## ACKNOWLEDGMENTS

V.V.C. and E.E.F. obtained partial financial assistance from CAPES, CNPq, and FAPESP. O.V.P. acknowledges support of the Computational Materials Sciences Program funded by the U.S. Department of Energy, Office of Science, Basic Energy Sciences, under Award Number DE-SC00014607.

## CONTACT INFORMATION

E-mail for correspondence: vvchaban@gmail.com (V.V.C.), fileti@gmail.com (E.E.F.), prezhdo@usc.edu (O.V.P.)